# $H_2S$ assisted contact engineering: a universal approach to enhance hole conduction in all TMD Field-Effect Transistors and achieve ambipolar CVD $MoS_2$ Transistors

Ansh[1], Jeevesh Kumar[1], Ravi K Mishra[2], Srinivasan Raghavan[2] and Mayank Shrivastava[1]

(1) Department of Electronic Systems Engineering, Indian Institute of Science, Bangalore-560012, Karnataka, India

(2) Centre for Nanoscience and Engineering (CeNSE), Indian Institute of Science, Bangalore-560012, Karnataka, India

Abstract

Unlike Si, 2-dimensional (2D) Transition Metal Dichalcogenides (TMDs) offer atomically thin channels for carrier transport in FETs. Despite advantages like superior gate control, steep sub-threshold swing and high carrier mobility offered by 2D FET channels, process related challenges like lack of selective doping techniques like implantation and CMOS compatible process for fabrication of 2D TMD based FETs hinder the anticipated viability of 2D semiconductor technology for future electronic applications. In this letter, we demonstrate a process oriented approach to realize superior ambipolarity in 2D FETs based on TMDs like Molybdenum disulfide ($MoS_2$), Tungsten disulfide ($WS_2$), Molybdenum diselenide ($MoSe_2$) and Tungsten diselenide ($WSe_2$) by enhancing hole current by multiple orders of magnitude in otherwise strong N-type transistors. The method involves Hydrogen Sulfide ($H_2S$ gas) assisted contact engineering of N-type FETs to introduce surface states that alter device behavior. Based on material characterization and bandstructure calculations, physical insights have been developed to understand the effect of such a contact engineering technique. Subsequently, this technique has been demonstrated to alter device behavior by enhancing hole conduction in originally N-type exfoliated ($MoS_2$, $WS_2$, $MoSe_2$ and $WSe_2$) and CVD grown ($MoS_2$) TMD samples to confirm its potential towards enabling the feasibility of 2D semiconductor device technology.

Since decades, Silicon (Si) has dominated the semiconductor industry due to its moderate bandgap and high carrier mobility. Apart from its favorable physical properties, realization of adequate process steps like etching, diffusion and implantation contributed towards technological viability of the Si based semiconductor industry for electronic applications. In order to match the rapidly growing demand for high performance and energy efficient electronic circuits, the Si industry advanced towards transistors as small as few nanometers in channel length, as suggested by Moore's law[1]. Consistent scaling of the device dimensions and limitations on the operating voltage ($V_{DD}$) has led to performance degradation of bulk-Si transistors due to Short Channel Effects (SCE), power density constraints and higher static power dissipation respectively. Mitigating SCEs in bulk-Si transistors has become a major challenge and efforts to marginalize and eliminate them were consistently made to achieve the desired performance as predicted by the Moore's law[2-9]. However, scaling beyond hundreds of nanometers starkly pushed the limits of Si transistors into the regime where quantum effects dominate leading to reemergence of SCEs that degrade performance. It was identified that the 3D nature of Si channel is not serving against SCEs at shorter channel lengths and superior gate control can be achieved by using either ultra-thin channel or multiple gates or both[10]. Hence, atomically thin Graphene and 2D TMDs [11-14] offer great potential to be used as channel material in short channel FETs. Unlike Graphene, atomically thin semiconducting TMDs like Molybdenum disulfide ($MoS_2$), Tungsten disulfide ($WS_2$), Molybdenum diselenide ($MoSe_2$) and Tungsten diselenide ($WSe_2$) are predicted to be suitable for complementary logic applications. The atomically thin conducting channel of 2D semiconductors restricts the use of implantation techniques to selectively dope and realize both electron and hole conduction. Techniques have been theoretically studied[15-17] and few have been experimentally realized to achieve desired polarity of the FETs. These techniques mostly include surface charge transfer by ionic sources [18-22] and few others involve non-selective doping during the material growth process [23, 24]. Such techniques primarily involve surface adsorption of ions as donors or acceptors through wet chemical treatment of the 2D semiconductor surface. Non-selectivity and incompatibility of wet chemical processes to CMOS fabrication process broaden the gap between 2D semiconductor research and technology development.

A decent progress on 2D TMD device research has been made in which $MoS_2$ and $WS_2$ FETs have been experimentally realized and are identified as N-type semiconductors irrespective of the contact metal [25-30]. Unlike the Sulfur based TMDs, $MoSe_2$ and $WSe_2$ are known to exhibit relatively enhanced ambipolar behavior [31-34] owing to their bandgap and bandstructure alignment with metal Fermi level. Technology development for a certain semiconductor requires realization of NMOS and PMOS transistors for which it is of prime importance that the semiconductor supports electron as well as hole conduction. However 2D TMDs like $MoS_2$ and $WS_2$, as discussed earlier, are n-type and support hole conduction only at large gate electric fields. Ambipolar behavior in $MoSe_2$ and $WSe_2$ FETs has been found to be dependent on thickness of the channel and contact metal and therefore it is difficult to realize carrier concentration of opposite polarity in monolayer TMDs due to large bandgap. In this work, a universal technique, compatible with the CMOS fabrication process, has been introduced that leads to enhanced hole conduction in $MoS_2$, $WS_2$, $MoSe_2$ and $WSe_2$ FETs. Physical insights into the surface chemistry and theoretical studies on bandstructure and density of states (DOS) led to a generic idea of introduction of Sulfur (S) atoms on the surface of TMDs to enhance hole injection at the contacts.

$H_2S$ is widely used as the precursor for S during the growth (CVD) process of S based TMDs like $MoS_2$ and $WS_2$ primarily because it decomposes on transition metal surface at high temperature. However, studies have shown low temperature partial decomposition of $H_2S$ on TMD that introduces extra S atoms on the surface followed by a complete catalytic decomposition to remove these S atoms [35]. Such a mechanism is feasible inside a TMD CVD growth reactor with $H_2S$ maintained at a certain pressure while the target is kept at a low temperature. The mechanism discussed in [35] has been illustrated in figure 1. At temperature as low as 350 °C, $H_2S$ gets adsorbed on the surface of TMD with the formation of $H_2S$-S bonds followed by release of $H_2$ thereby leaving behind S atoms on the surface of the target (TMD).

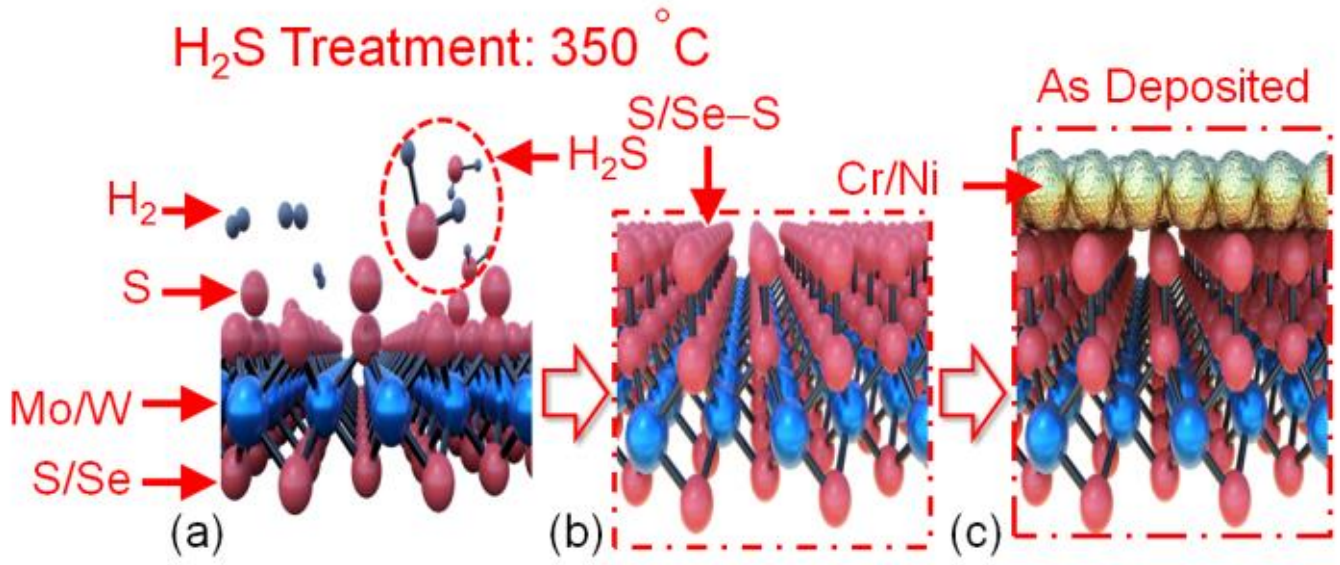

Figure 1: Low temperature partial decomposition of $H_2S$ on TMD surface as explained by Startsev et. al.[35]. (a) TMD samples kept inside a CVD growth furnace at 350 C and 20 torr partial pressure. (b) $H_2S$ gets adsorbed on the surface forming S-S bonds and subsequent release of $H_2$. (c) Contact metal deposited on top of $H_2S$ exposed regions for Source/Drain (S/D) electrodes.

TMD samples were exposed to $H_2S$ at 350 °C inside a CVD growth furnace. From [35], it is expected that S gets introduced on the surface of TMD films, however, the same needs to be experimentally verified. Moreover, such a process, if not controlled, on thin TMD films can lead to damage to the crystal lattice and eventually alter the fundamental molecular structure of the film. In order to verify that the fundamental molecular structure is held intact even after exposure to $H_2S$, Raman spectra were captured for all TMDs pre and post exposure and compared. It is observed in figure 2 that the signature Raman peaks ($E_{2g}$ and $A_{1g}$) for all TMDs ($MoS_2$, $WS_2$, $MoSe_2$ and $WSe_2$) are present in the Raman spectra before and after $H_2S$ exposure implying that samples retain their molecular structure post exposure. A blue shift is observed in the position of Raman peaks after $H_2S$ exposure which implies enhanced electron-phonon coupling[36]. Enhanced coupling is attributed to stronger interaction between electrons and surface added S (atoms after $H_2S$ exposure) owing to smaller size and high electronegativity of S atom. Moreover, reduced shoulders near the peaks are attributed to reduced defect density on the surface, post-exposure.

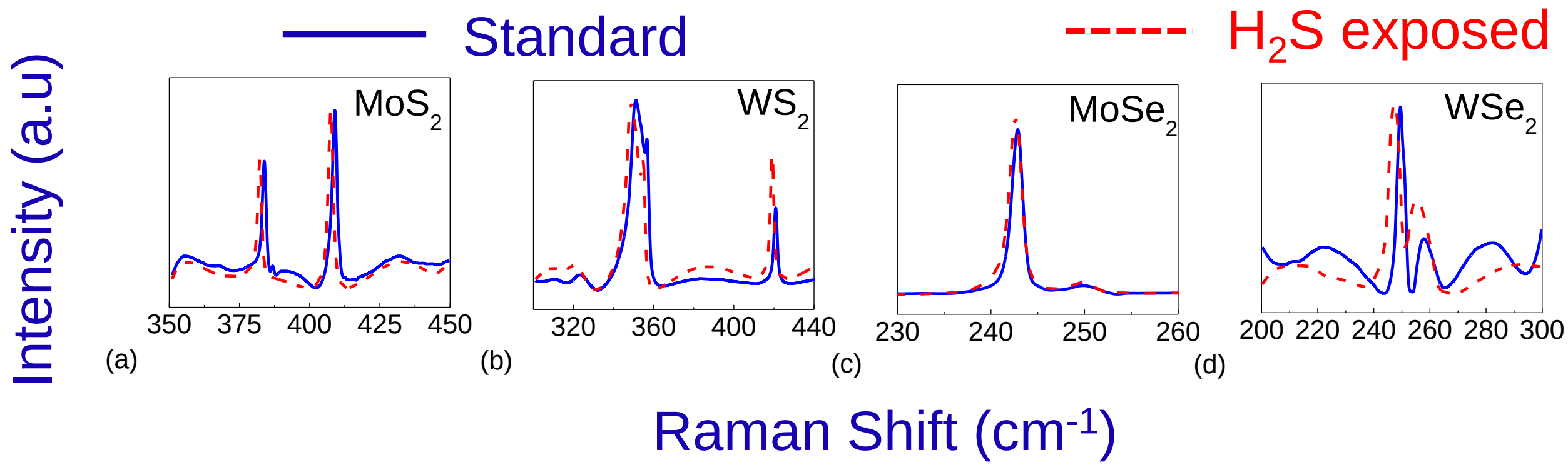

Figure 2: Comparison of Raman spectra of (a) $MoS_2$, (b) $WS_2$, (c) $MoSe_2$ and (d) $WSe_2$ before and after $H_2S$ exposure to investigate the effect of exposure on the surface of TMDs.

In order to investigate chemical changes on the surface of TMDs, X-ray Photoelectron Spectroscopy (XPS) spectra of unexposed and exposed samples were captured and compared. As shown in figure 3, S peaks are observed only in the XPS spectra of unexposed $MoS_2$ and $WS_2$ samples and not in that of $MoSe_2$. Interestingly, all samples ($MoS_2$, $WS_2$, $MoSe_2$ and $WSe_2$) exhibit S peaks in their XPS spectra post-exposure. It is observed that, post-$H_2S$ exposure, intensity of peaks corresponding to 2p orbitals from S atoms increased in $MoS_2$ and $WS_2$ samples whereas S-2p peaks appeared in XPS spectra of $MoSe_2$ and $WSe_2$ which are otherwise absent in XPS spectra of pristine $MoSe_2$ and $WSe_2$ samples. It is therefore validated that S atoms are incorporated on the surface of TMDs after $H_2S$ exposure. As inferred from Raman spectra, lowering of shoulder around prominent peaks in XPS spectra also implies that the defect density in these samples has been significantly reduced as a result of $H_2S$ exposure.

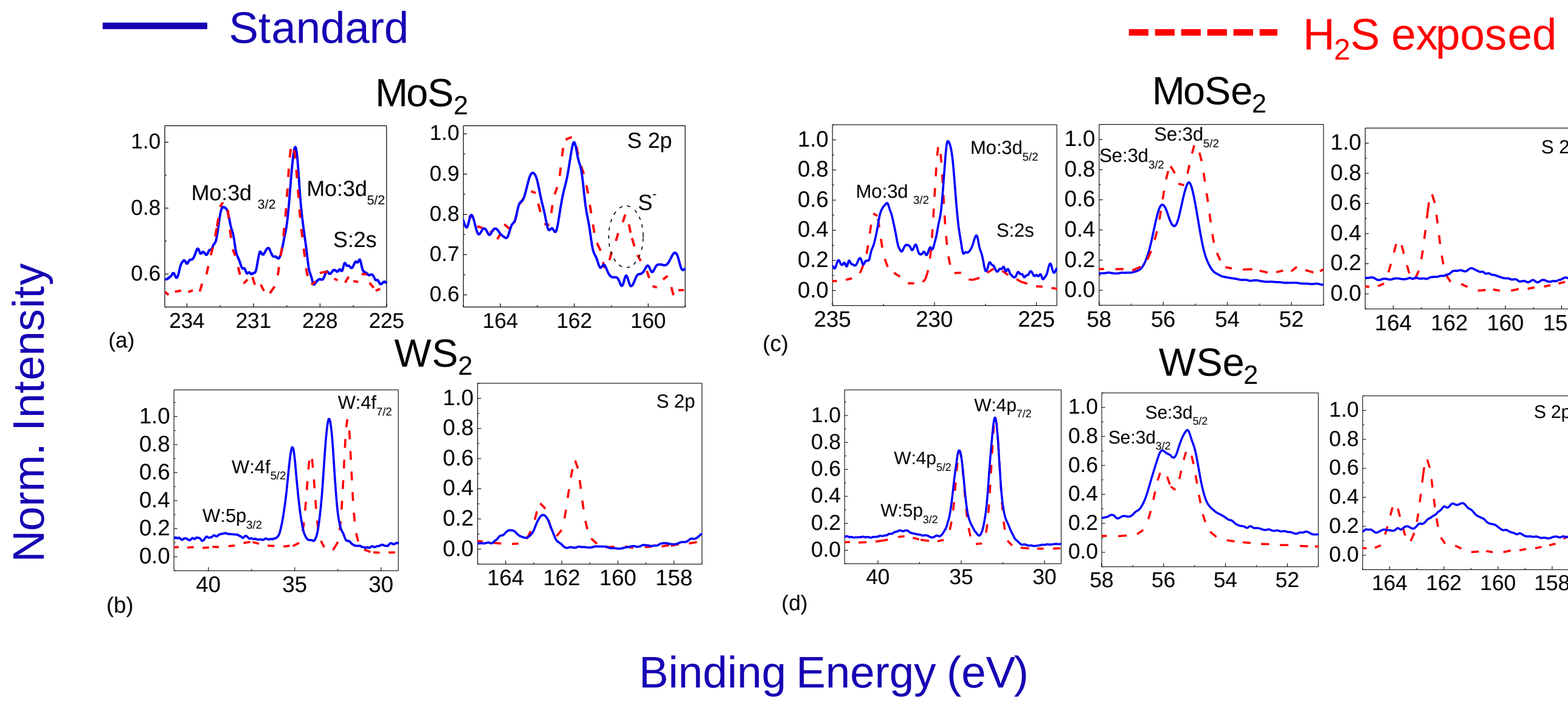

Figure 3: Comparison of XPS spectra before and after $H_2S$ exposure for (a) $MoS_2$, (b) $WS_2$, (c) $MoSe_2$ and (d) $WSe_2$. Additional S peaks are observed in Se based TMD samples along with reduced shoulders around prominent peaks. This validates the presence of S atoms on the surface and reduced defect density.

It is interesting to investigate the effect of surface S atoms on the behavior of TMD based FETs. In order to fabricate TMD FETs, TMD crystals are first exfoliated using the standard scotch tape method on thermally grown 90 nm $SiO_2$ on p-type Si samples. Such a stack is used to fabricate back-gated FETs where p-type Si is used as the back gate. After exfoliation, few layered flakes are identified through optical microscope. In order to eliminate variation due to flake thickness, standard (unexposed to $H_2S$) and contact engineered (exposed to $H_2S$) FETs are fabricated on a single flake with uniform thickness along its length. Flake identification was followed by patterning channel region through e-beam lithography so that it can be masked with $Al_2O_3$, which is deposited using e-beam evaporator, to prevent $H_2S$ exposure of channel. Standard FETs are fabricated by patterning source/drain contacts through e-beam lithography followed by metal deposition and lift-off. Post-lift-off, samples are annealed at 200 ˚C for 10 minutes to ensure proper contact between metal and TMD. The fabricated standard FETs with $Al_2O_3$ passivation on channel are electrically characterized. Immediately after DC characterization, samples are exposed to $H_2S$ at 350 ˚C inside a CVD growth chamber followed by fabrication of contact engineered FET on the left over region of the same flake on which standard FETs are fabricated. The fabrication process of contact engineered FETs involved S/D contact patterning and metal deposition followed by lift-off and post-anneal at 200 ˚C. The channel of contact engineered FETs was already masked as a result of the $Al_2O_3$ deposition step discussed earlier. It is important to point out that the fabrication process followed for standard and contact engineered FETs is exactly the same without any process variation. This is done to ensure a legitimate comparison between standard and contact engineered FETs so that the actual effect of S atoms on the surface can be probed. Figure 4 shows the schematic of standard and contact engineered FETs fabricated on the same TMD film.

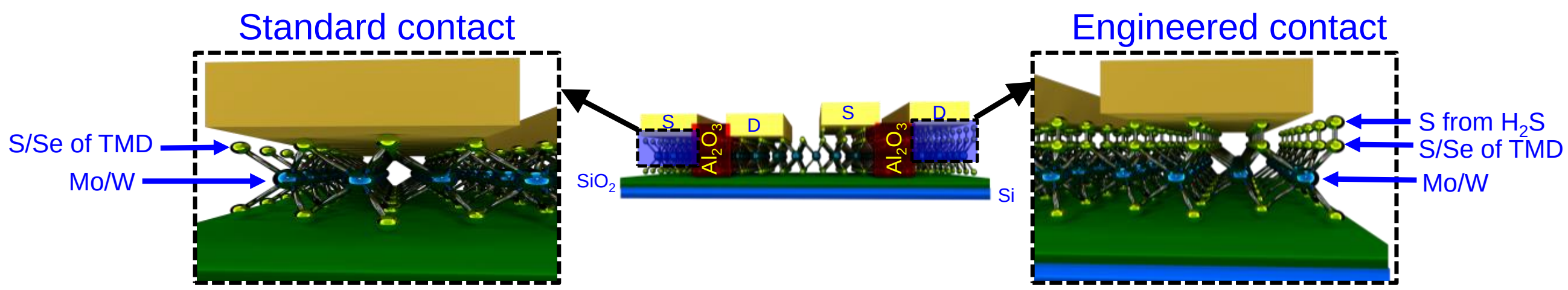

Figure 4: Scheme followed to fabricate standard and contact engineered FETs on the same TMD film on top of $SiO_2$/Si sample with $Al_2O_3$ passivation on the channel region.

Transfer characteristics of standard and contact engineered FETs of all TMDs are compared in figure 5. It can be seen that standard FETs based on TMDs with Se as the Chalcogen constituent exhibit strong ambipolar behavior than that on S based TMDs. This is consistent with other works and is attributed to relatively lower bandgap in $MoSe_2$[33] and Fermi level pinning (FLP) at lower energy in $WSe_2$[31]. Moreover, it is worth mentioning that all standard FETs based on four different TMDs exhibit dominant n-type behavior, consistent with the literature for the contact metals in use. Standard $MoS_2$, $WS_2$ and $MoSe_2$ FETs exhibit n-type behavior with a threshold close to $V_{GS}$ = -15 V whereas for standard $WSe_2$ FETs, threshold is close to $V_{GS}$ = -20 V. $H_2S$ exposed FETs at the contacts (contact engineered) exhibit enhanced hole as well as electron conduction as shown in figure 5. It is observed that, as the gate voltage is swept to negative values beyond $V_{GS}$ = 0 V, FETs reach OFF state and only a small leakage (OFF state) current flows through the channel. Beyond the point when $V_{GS}$ = -15 V, all contact engineered FETs start exhibiting significant current across the channel which was absent/marginal in standard FETs. This is attributed to significant reverse band bending that leads to formation of hole-Schottky barrier and hence hole injection at relatively higher negative gate voltage than in standard FETs. From transfer characteristics of all FETs it can be inferred that $H_2S$ assisted contact engineering causes some change in the surface that leads to enhanced hole injection. Another observation is that contact engineered FETs have higher threshold voltage ($V_{TH}$) than standard FETs. This can be viewed as a doping phenomenon where excess holes are incorporated in the crystal lattice. However, enhanced electron as well as hole current after contact engineering contradicts this hypothesis of doping as excess hole concentration is not expected to simultaneously improve electron current. Improved electron current is attributed to reduced tunneling barrier as a result of stronger bonding at the metal-$H_2S$ exposed TMD surface.

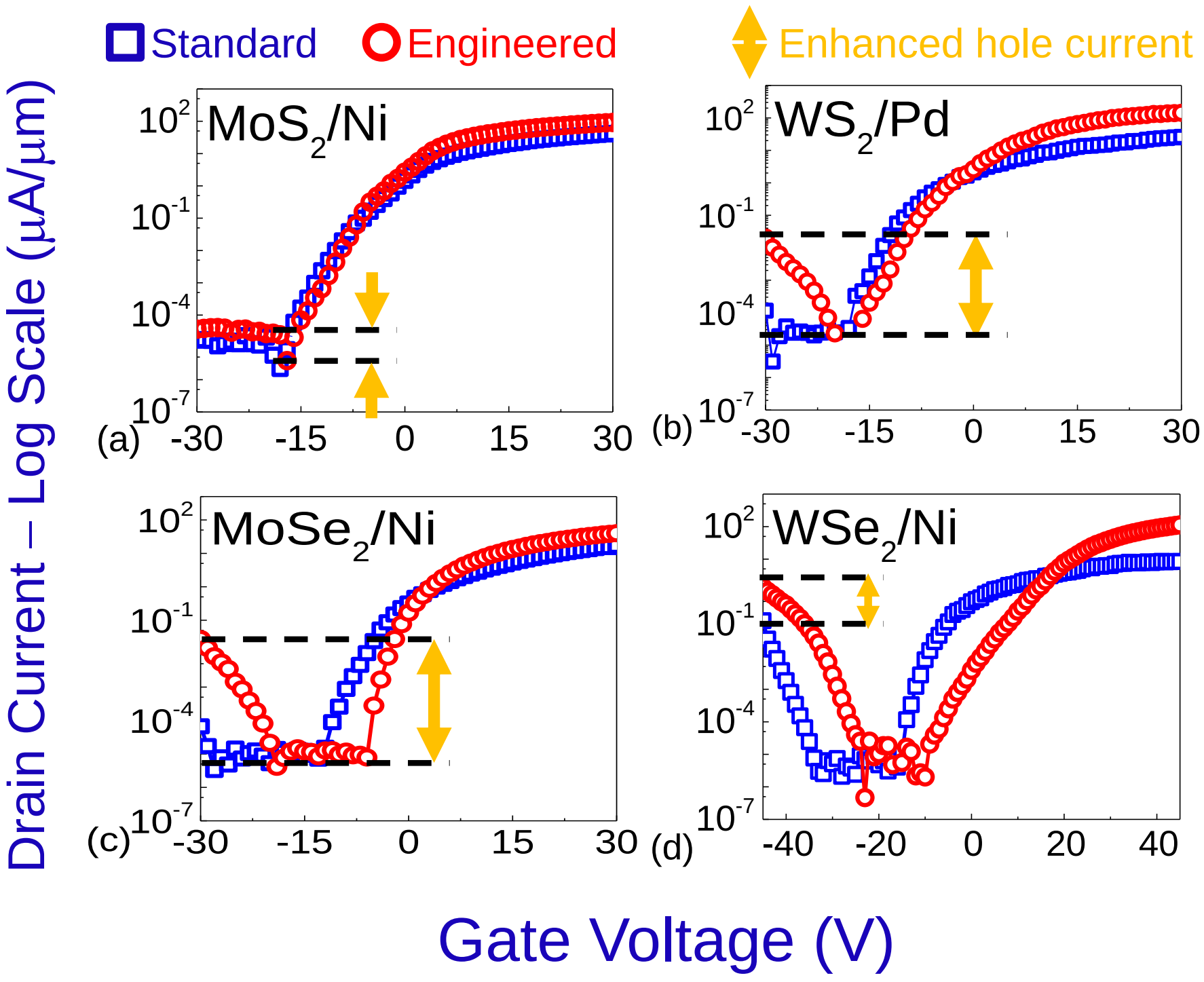

Figure 5: Transfer characteristics of standard and contact engineered FETs on few layer exfoliated (a) $MoS_2$, (b) $WS_2$, (c) $MoSe_2$ and (d) $WSe_2$. In general, hole conduction is found to have enhanced across all the TMDs post-$H_2S$ assisted contact engineering. Along with improved hole current, electron current has improved for all the TMDs which is attributed to improved bonding at the contact due to the presence of S atoms on the surface.

When compared with FETs on few layer exfoliated $MoS_2$ channel, FETs on monolayer CVD $MoS_2$ channel are expected to exhibit stronger unipolar behavior owing to a large band gap and high defect density that pins the Fermi-level much closer to CBM[27]. Therefore, the challenge is to realize ambipolar FETs on monolayer CVD $MoS_2$. FETs are fabricated on full-coverage CVD monolayer $MoS_2$ (from 2D semiconductors) grown on 90 nm thick $SiO_2$ on p-type Si sample. The sample is patterned to etch (inside an RIE chamber in $O_2$ plasma) undesired $MoS_2$ regions so that individual $MoS_2$ channels can be isolated. $Al_2O_3$ is then deposited to mask the channel region against $H_2S$ exposure followed by S/D patterning and metal deposition for fabrication of standard FETs and subsequent DC characterization. Next, the samples are exposed with $H_2S$ at desired chamber conditions for the proposed contact engineering followed by patterning and metal deposition for S/D electrodes of contact engineered FETs. Figure 6, shows transfer characteristics of CVD monolayer $MoS_2$ FET with Ni contacts. It is observed that standard FETs exhibit unipolar N-type behavior with $V_{TH}$ = -20 V and a drain current modulation of ~ $10^7$. Owing to large defect density, large bandgap and large work function difference, CVD monolayer $MoS_2$ FETs with Ni are expected to offer low performance as compared to their exfoliated few-layered counterparts with low work function metal contacts. Interestingly, the strong unipolar behavior of standard FETs on CVD monolayer $MoS_2$ disappeared after contact engineering. Contact engineered FETs exhibit enhanced hole current than standard FETs by 3 orders of magnitude. On the other hand, huge positive shift in $V_{TH}$ led to reduced electron current by a factor of 2 at $V_{GS}$ = 45 V. In general, it is observed that, contact engineered FETs on CVD monolayer $MoS_2$ offer comparable hole and electron conduction unlike standard FETs which are unipolar.

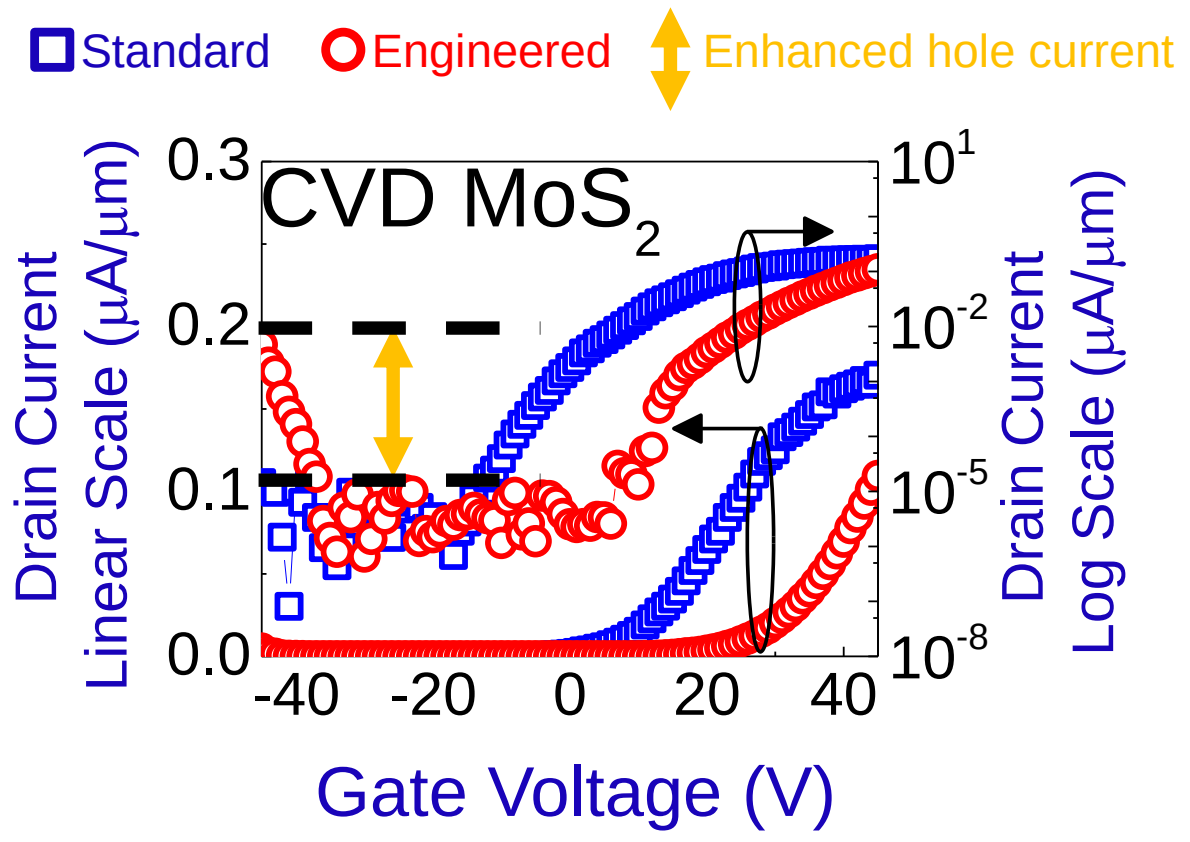

Figure 6: Transfer characteristics of standard and contact engineered FETs on CVD monolayer $MoS_2$. $H_2S$ assisted contact engineering has led to enhanced hole conduction through the transistor thereby achieving enhanced ambipolarity in CVD monolayer $MoS_2$ based FETs.

In order to explain the observed contact engineering led enhanced hole injection and positive shift in $V_{TH}$, we performed DFT simulations to calculate the bandstructure of various surface topologies of all four TMDs. Three different surface topologies- (a) defect-free, (b) surface with Chalcogen vacancy and (c) defect-free surface with S at an interstitial site are simulated to extract the bandstructure. In general, it is observed that unlike a surface with chalcogen vacancy where a donor state is present near conduction band minimum (CBM), S at an interstitial site results in acceptor states near Valence band maximum (VBM). Such introduction of surface states near VBM has been observed consistently for all TMDs- $MoS_2$, $WS_2$, $MoSe_2$ and $WSe_2$ in figures 7, 8, 9 and 10 respectively. It is well known that surface states lead to FLP of metal Fermi level close to CBM or VBM depending on their location in the bandgap[37, 38]. Therefore, we propose that $H_2S$ assisted contact engineering leads to S atoms on the surface (validated by XPS spectra in figure 3) that in turn introduce surface acceptor states thereby pinning the metal Fermi level close to VBM of all TMDs. As a result of FLP near VBM, enhanced hole injection is achieved along with positive shift in $V_{TH}$.

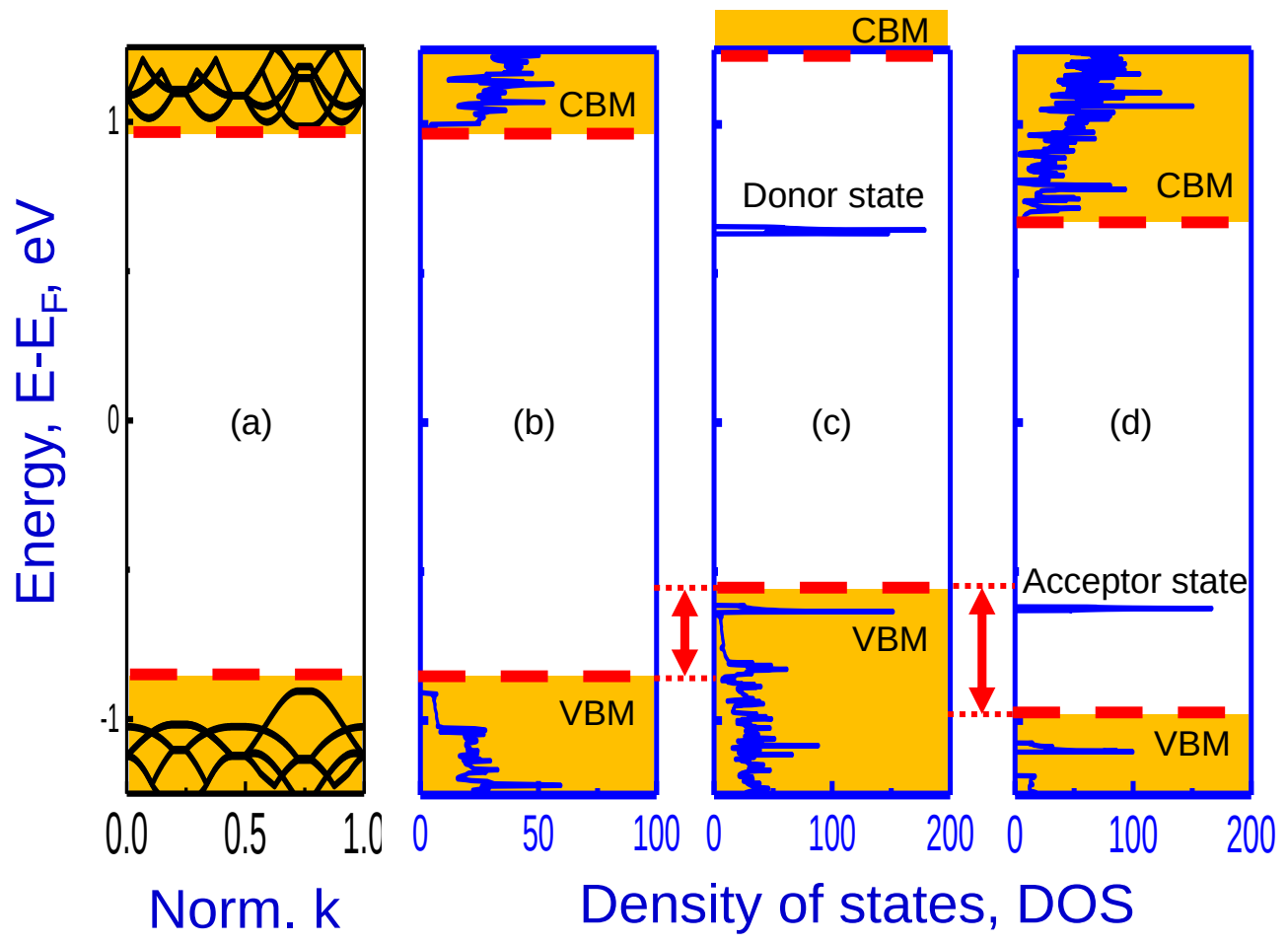

Figure 7: (a) Bandstructure of monolayer $MoS_2$. (b) Density of States (DOS) in defect-free monolayer $MoS_2$. Absence of bandgap states is intuitive owing to lack of surface states in defect-free $MoS_2$. (c) DOS in monolayer $MoS_2$ with a S vacancy. A S-

vacancy introduces a donor state near CBM with a significantly high DOS as shown here. (d) DOS in monolayer $MoS_2$ with extra S at an interstitial site that leads to an acceptor state close to VBM with finite DOS.

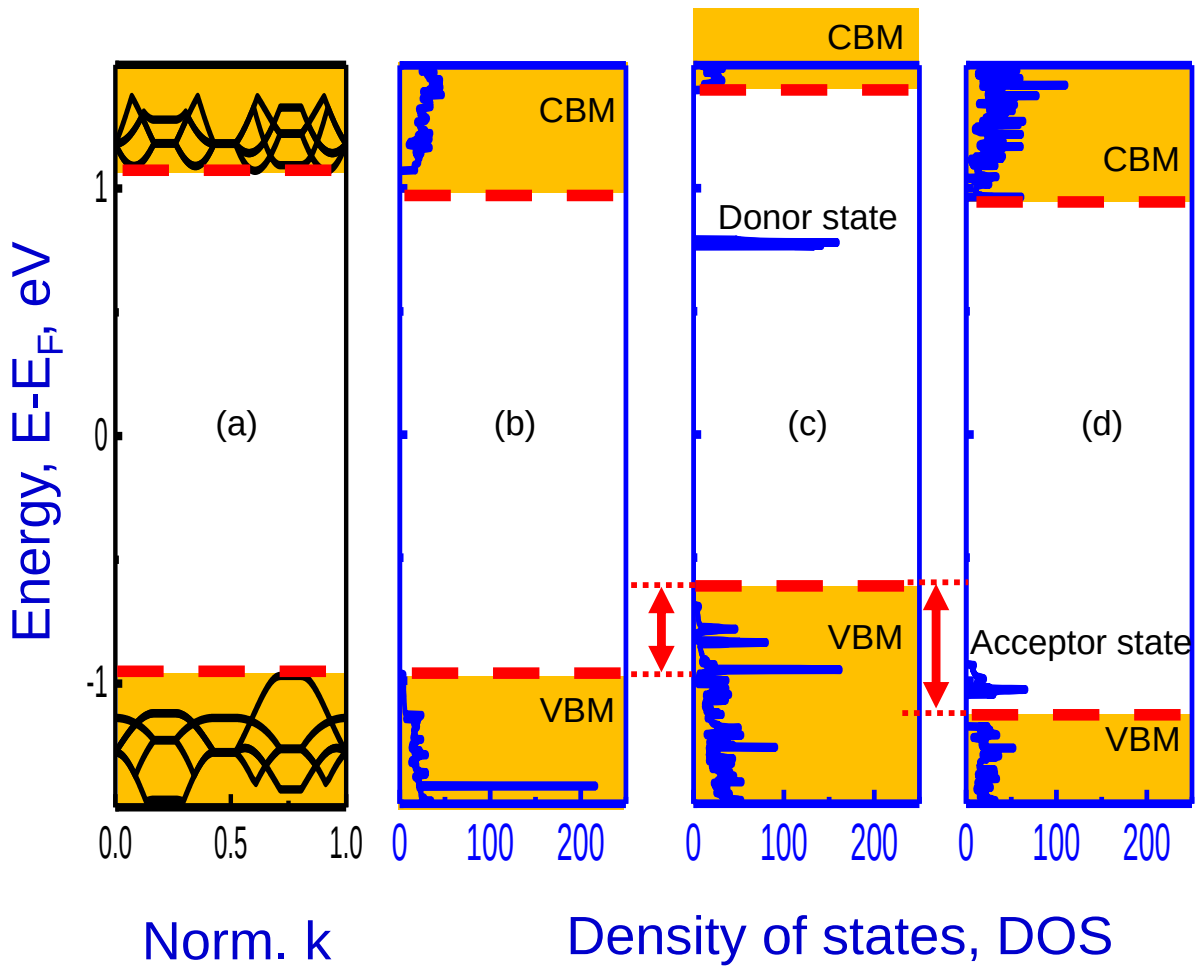

Figure 8: (a) Bandstructure of monolayer $WS_2$. (b) Density of States (DOS) in defect-free monolayer $WS_2$. Absence of bandgap states is intuitive owing to lack of surface states in defect-free $WS_2$. (c) DOS in monolayer $WS_2$ with a S vacancy. A S- vacancy introduces a donor state near CBM with a significantly high DOS as shown here. (d) DOS in monolayer $WS_2$ with extra S at an interstitial site that leads to an acceptor state close to VBM with finite DOS.

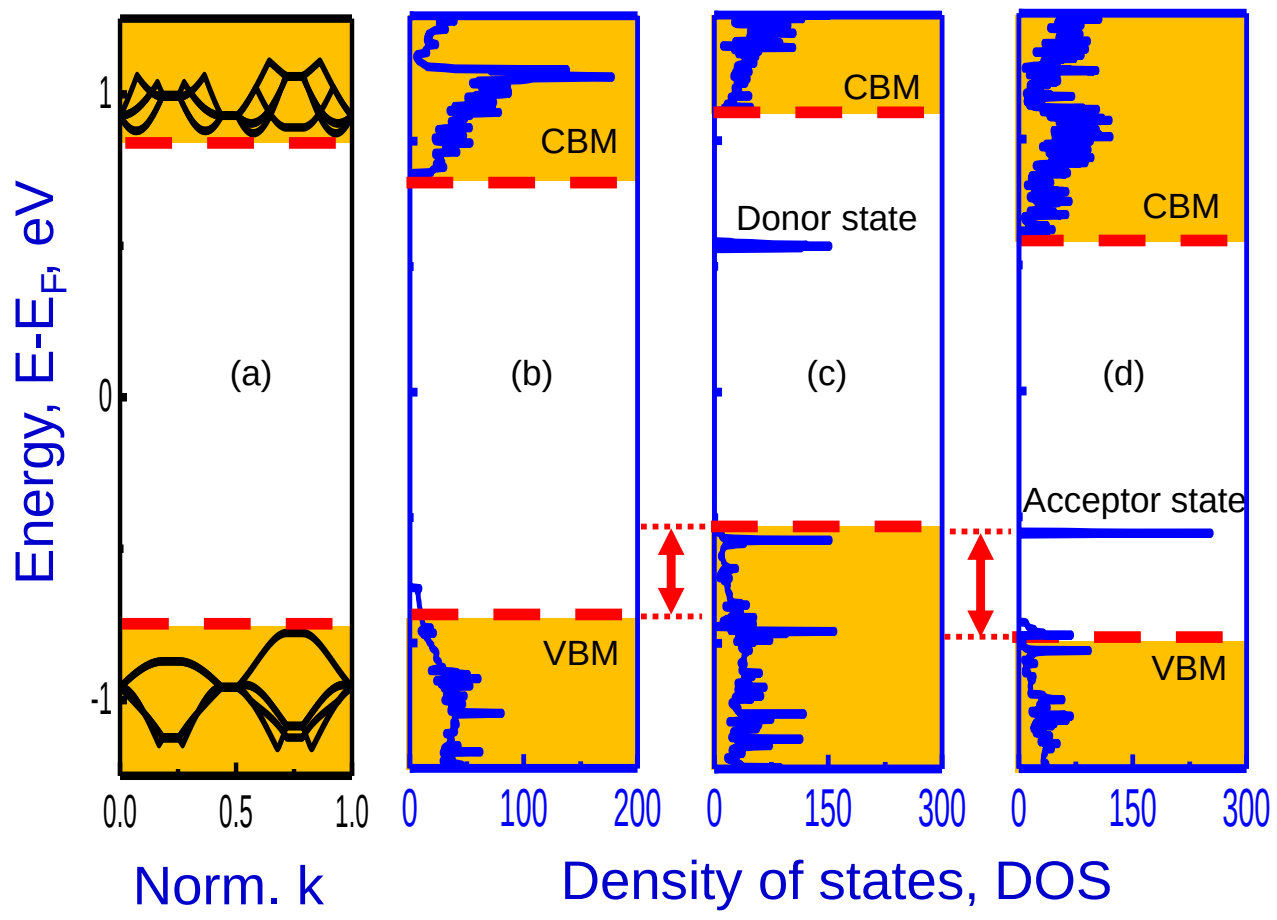

Figure 9: (a) Bandstructure of monolayer $MoSe_2$. (b) Density of States (DOS) in defect-free monolayer $MoSe_2$. Absence of bandgap states is intuitive owing to lack of surface states in defect-free $MoSe_2$. (c) DOS in monolayer $MoSe_2$with a Se vacancy. A Se- vacancy introduces a donor state near CBM with a significantly high DOS as shown here. (d) DOS in monolayer $MoSe_2$with extra S at an interstitial site that leads to an acceptor state close to VBM with finite DOS.

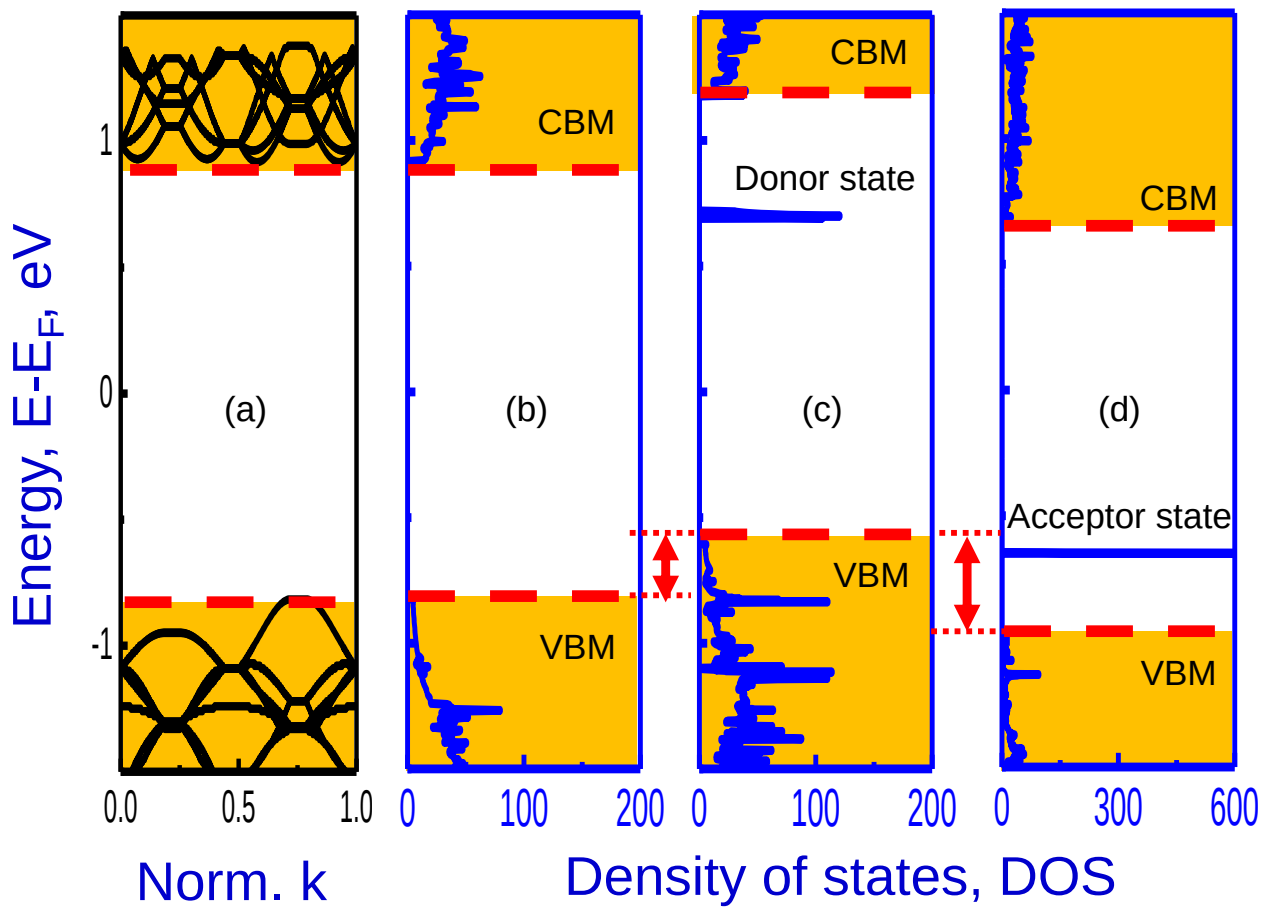

Figure 10: (a) Bandstructure of monolayer $WSe_2$. (b) Density of States (DOS) in defect-free monolayer $WSe_2$. Absence of bandgap states is intuitive owing to lack of surface states in defect-free $WSe_2$. (c) DOS in monolayer $WSe_2$with a Se vacancy. A Se- vacancy introduces a donor state near CBM with a significantly high DOS as shown here. (d) DOS in monolayer $WSe_2$with extra S at an interstitial site that leads to an acceptor state close to VBM with finite DOS.

Physical insights into the effect of $H_2S$ exposure on the surface of TMDs and subsequent impact on device performance can be used to explain the FET switching behavior through band theory. As explained in figure 11, in standard FETs FLP occurs close to CBM that results in strong N-type behavior. A positive gate voltage sweep bends the bands below due to which the Schottky barrier width (SBW) reduces thereby improving electron tunneling current and and overall increase in current through the device for higher positive gate voltages. A negative voltage sweep bends the bands upwards and subsequently at sufficiently negative voltage Schottky barrier for hole injection is realized. Beyond this voltage, the FET enters a hole conduction regime. In contact engineered FETs, FLP occurs at lower energies, i.e. close to VBM (as discussed earlier) and therefore the Schottky barrier for hole injection is realized at a more positive voltage than that in case of standard FETs. Therefore, enhanced hole conduction is achieved through contact engineered FETs for the same gate voltage range as in case of standard FETs, shown in figure 12.

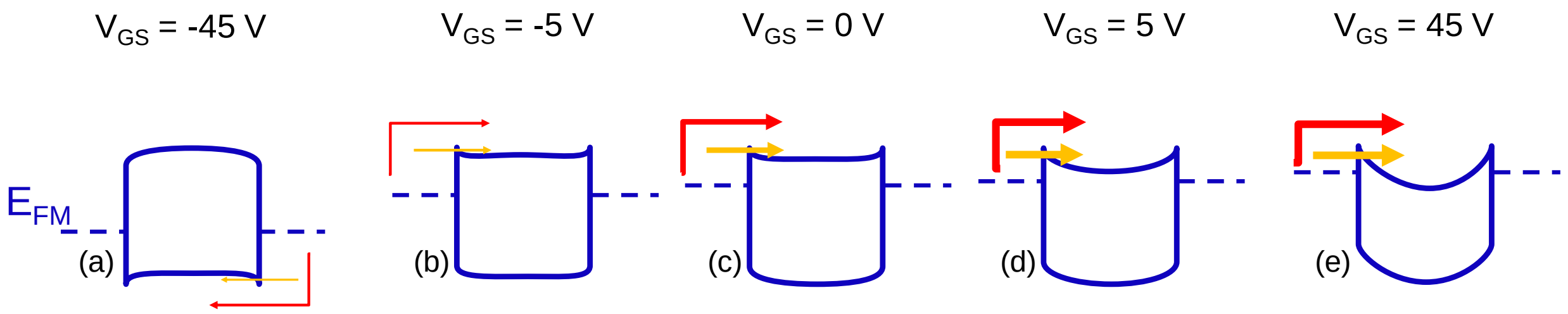

Figure 11: Switching mechanism of standard Schottky barrier FETs. (a) and (b): Marginal hole conduction and electron conduction, respectively, at negative gate voltage owing to FLP close to CBM corresponding to the OFF state of the device. (c) Significant electron current at $V_{GS}$ = 0 V implying normally ON behavior of the transistor. (d) and (e): Enhanced electron conduction for positive gate voltage due to downward band bending causing narrower Schottky barrier to facilitate more electron injection across the contact, corresponding to ON state of the transistor. Note: Yellow and red arrows correspond to tunneling and thermionic components of the total current. Tunneling electron current increases exponentially with increasing gate voltage owing to reduced SBW.

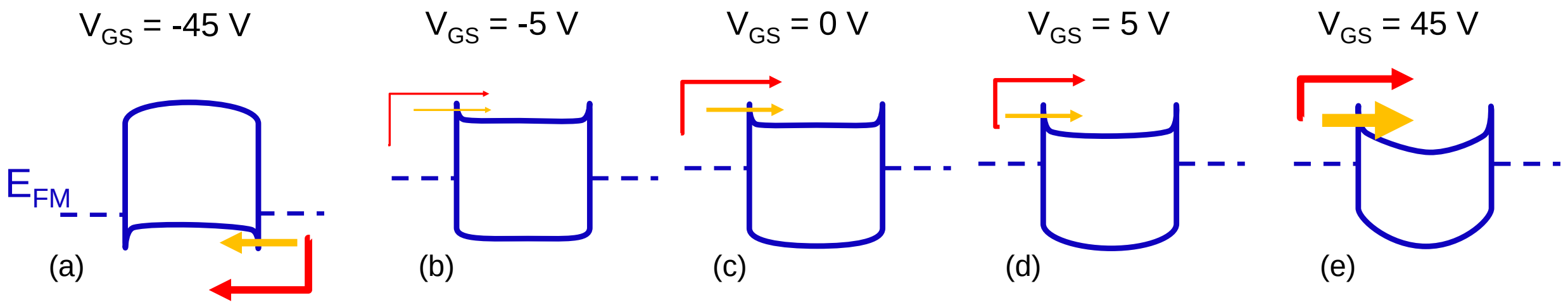

Figure 12: Switching mechanism of contact engineered Schottky barrier FETs. (a): enhanced hole conduction, as compared to standard FETs, at negative gate voltage owing to FLP close to VBM corresponding to the ON state of the device in the hole conduction regime. (b): device in OFF state with marginal electron current due to large SBH caused by FLP closer to VBM as compared to that in standard FET. (c) Significant electron current at $V_{GS}$ = 0 V implying normally ON behavior of the transistor. (d) and (e): Enhanced electron conduction for positive gate voltage due to downward band bending causing narrower Schottky barrier to facilitate more electron injection across the contact, corresponding to ON state of the transistor. Enhanced electron current is also observed in contact engineered transistors which is attributed to improved bonding at the contacts which narrows the tunneling barrier thereby enhancing tunneling current. Note: Yellow and red arrows correspond to tunneling and thermionic components of the total current. Tunneling electron current increases exponentially with increasing gate voltage owing to reduced SBW.

Such a Chalcogen based technique to alter transistor behavior and controllably achieve enhanced hole current and subsequent improvement in ambipolarity has been demonstrated for the first time. Besides involving dry chemistry, this technique is CMOS process compatible and hence scalable. It also supports bulk processing as $H_2S$ exposure of multiple samples can be performed at the same time inside a single chamber. As discussed earlier, in order to establish 2D semiconductor technology, fabrication methods need to be developed to achieve both NMOS and PMOS transistors so that challenges in realizing complementary logic application can be addressed. This technique promises great potential in realizing PMOS transistors for all TMDs and therefore it is projected to be a generic method to achieve 2D TMD based PMOS FETs.